\documentclass{camera}
\usepackage{graphicx}  % uncomment this if your want to insert figures

\begin{document}

%%%%%%%%%%%%%%%%%%%%%%%%%%%%%%%%%%%%%%%%%%%%%%%%%%%%%%%%
% The title, only the first letter capitalized; if you want to split it in
% two or more lines, put a \\ macro at each line break
% example: 

\title{Early time optical polarization of GRB Afterglows:\\GRB 060418 and GRB 090102 }

%%%%%%%%%%%%%%%%%%%%%%%%%%%%%%%%%%%%%%%%%%%%%%%%%%%%%%%%
% The author(s), separated by commas; do not put a
% comma before the last author, use instead the \and
% macro which produces a normal ``and'' in the
% caps/small caps context
%
\author{I. A. Steele$^1$, C. G. Mundell$^1$, R. J. Smith$^1$, S. Kobayashi$^1$ \\ \and C. Guidorzi$^2$}

%%%%%%%%%%%%%%%%%%%%%%%%%%%%%%%%%%%%%%%%%%%%%%%%%%%%%%%%
%
\organization{$^1$Astrophysics Research Institute, Liverpool JMU \\ $^2$
  Physics Department, University of Ferarra}

\maketitle

\begin{abstract}
 
RINGO on the Liverpool Telescope has now measured the optical polarization of GRB
060418 (where a 2$\sigma$ upper limit of $P<8$ \% was determined) and
GRB 090102 (when a detection of $P ~ 10 \pm 1$ \% was made).
We discuss the implications of these observations for the various
competing models of GRB jet magnetization and describe a possible
unified model that can explain both measurements.

\end{abstract}

%%%%%%%%%%%%%%%%%%%%%%%%%%%%%%%%%%%%%%%%%%%%%%%%%%%%%%%%
% Write the text starting from here and using the usual
% LaTeX commands.
%

If GRB ejecta have large scale magnetic fields, then the prompt $\gamma-$ray,
X-ray and optical emission and the reverse shock emission should be
polarized\cite{lyu06}.  Optical polarization is generally measured via a ratio of
fluxes in two or more polarizations.  Traditionally this is done by
using a modulator/analyzer pair taking separate
exposures.  However this approach does not work for a rapidly varying
object such as a GRB afterglow - a one per-cent ``fade'' between
subsequent exposures implying a false one per-cent polarization signal
for example.  It was therefore decided to develop an instrument for
the Liverpool Telescope\cite{lt} that can make a ``single-shot''
polarization measurement.  The instrument also required a wide
field of view (so that it could respond to SWIFT BAT alerts and be
on target during the prompt/reverse shock emission phases).  We
therefore developed and built the RINGO\cite{ringo} polarimeter, which
uses a fast rotating Polaroid to modulate the incoming beam, followed by 
corotating deviating optics that transfer each star image into a ring that is recorded on a CCD (Figure 1).  
Any polarization signal present in the incoming light is mapped out around the
ring in a $sin 2\theta$ pattern.  

\begin{figure}
\includegraphics[scale=0.5]{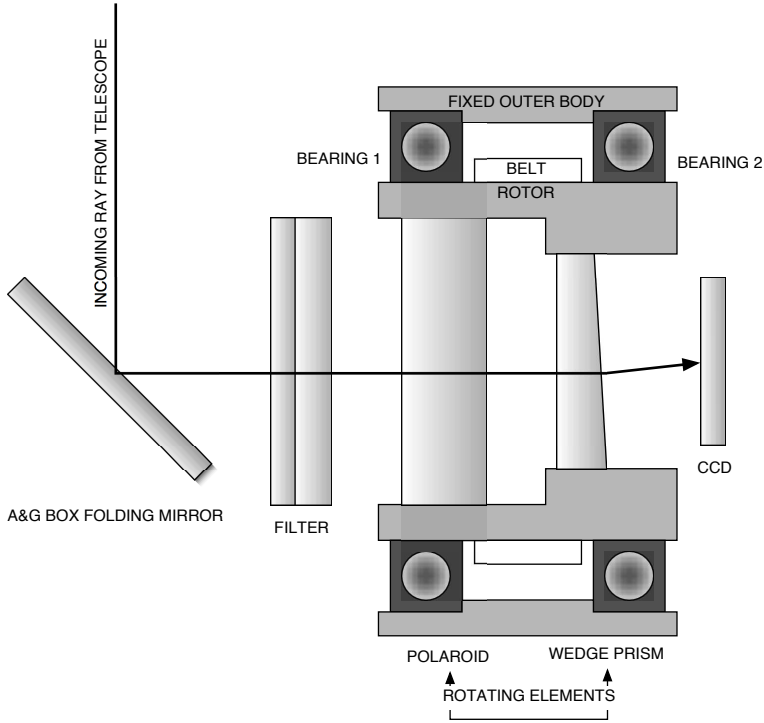}
\caption{Layout of the RINGO instrument.}
\label{fig01} % optional figure label, must be unique
\end{figure}

RINGO was first used in 2006, when it observed GRB~060418 at 203s after the gamma ray burst and coincident with the
 time of deceleration of the fireball.  At this time the reverse (assuming it was present) and forward shock 
components would have contributed equally to the observed light.  For GRB~060418 a 2$\sigma$ upper limit on optical 
polarization of P$<$8\% was measured in the combined light from the emitting regions\cite{mun07a}. 

RINGO was next used in 2009, when it observed
GRB~090102\cite{man09}.  The steep-shallow decay\cite{man09,kl09,cov09} of optical emission from GRB~090102 is characteristic 
of an afterglow whose early-time  light is dominated by  fading radiation generated in the reverse shock\cite{zkm03,gom08}.
A single 60-second RINGO exposure was obtained starting 160.8 seconds
after the trigger time. RINGO measured the optical (4600 - 7200{\AA}) polarization of GRB 090102 as $P=10.2\pm1.3$\%\cite{nature}  (Figure 2).

\begin{figure}
\includegraphics[scale=0.5, angle=270]{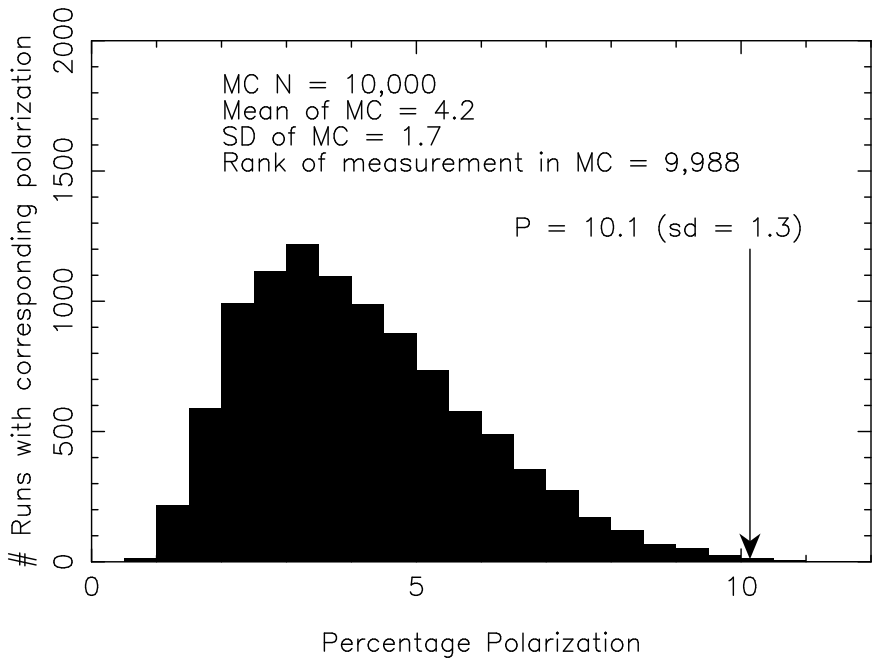}
\caption{Monte Carlo simulation demonstrating the significance of the GRB 090102 result. See \cite{nature} for details.}
\label{fig02} % optional figure label, must be unique
\end{figure}

In interpreting this measurement first we consider whether such a polarization could be produced via the 
production of magnetic instabilities in the shock front  (Figure 3(c)).  A very optimistic estimate of the coherence 
length can be made by assuming it grows at about the speed of light in the local fluid frame after the field is 
generated at the shock front - in this situation polarized radiation would come from a number of independent 
ordered magnetic field patches.  A measured polarization of 10\% is at the very uppermost bound for such a model
\cite{gw99} and therefore seems unlikely. As an alternative to the ``patch'' model, we have also considered the 
case where the observer's line of sight is close to the jet edge\cite{gru99} (Figure 3(b)).  In this case since
the magnetic fields parallel and perpendicular to the shock front could have significantly different averaged 
strengths\cite{ml99} a polarization signal can also be produced.  However applying this model to GRB~090102 we 
would have expected a steepening of the light curve (a ``jet-break'') just after the time of our polarization 
measurement rather than the observed flattening.  Similarly there is no evidence of a jet break in the X-ray light curve up to late times.  Furthermore, our detection of 10\% is much higher than the reported polarization signal of a few $\%$ associated with a jet break in late time afterglow of other events\cite{cov99, wij99}.  We also rule out an Inverse Compton origin for the optical polarization - a mechanism suggested to explain earlier $\gamma$-ray polarization measurements\cite{laz04} - in which lower energy photons are scattered to higher energies by colliding with electrons in the relativistic flow. If Inverse Compton emission is present, it is more likely to contribute to the high-energy X-ray and $\gamma$-ray bands than the optical band and again requires the observer's line of sight to be close to the edge of the jet (Figure 3(b)) to produce significant polarization which (as we have already discussed) is not the case for GRB~090102.

\begin{figure}
\includegraphics[scale=0.65]{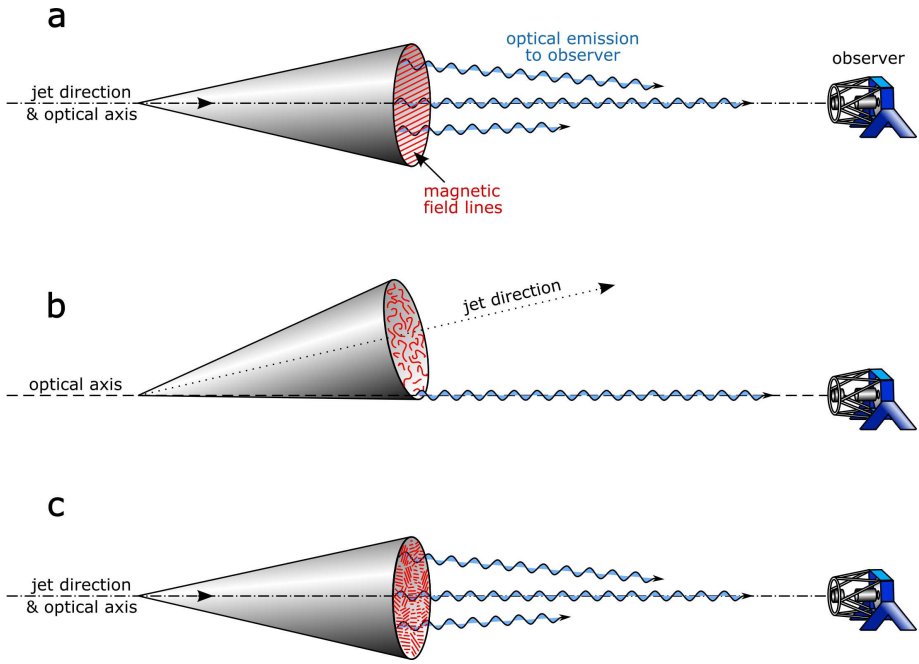}
\caption{Alternative models for GRB polarization}
\label{fig03} % optional figure label, must be unique
\end{figure}

It therefore seems apparent that in the case of GRB~090102 the high polarization signal requires the presence 
of large-scale ordered magnetic fields in the relativistic outflow (Figure 3(a)). As the measurement was 
obtained while the reverse-shock emission was dominant in GRB~090102, the detection of significant polarization 
provides the first direct evidence that such magnetic fields are present when significant reverse shock emission is 
produced.

Magnetization of the outflow can be expressed as a ratio of magnetic to kinetic energy flux, $\sigma$.  The degree of magnetization cannot
be sufficient for the jet to be completely Poynting flux dominated ($\sigma>1$)
since this would be expected to suppress a reverse shock\cite{mga09}.
We can therefore reconcile the
detection of polarization in GRB~090102 and the non-detection in GRB~060418 
in a unified manner if GRB jets have magnetization 
of $\sigma \sim 1$. In the GRB~060418 case, the jet would have had slightly higher magnetization than unity, 
resulting in the suppression of a reverse shock, while GRB~090102
would have $\sigma$ slightly smaller than unity, which is optimal to 
produce bright reverse shock emission.  Of course due to the small sample (only two 
objects), we can not rule out a possibility that each GRB jet had very
different magnetization.

Finally we note recent claims of rapidly ($\sim10$ s) 
variable $\gamma$-ray polarization from less than 4\% to 43\% ($\pm$25\%) in the prompt emission of GRB~041219A
\cite{gotz09}.  This lends further support to models with magnetized outflows and offers the possibility that the peak 
optical polarization from GRB~0901012
could have been even higher than that measured in our 60 second
exposure.  We are have therefore developed a new instrument (RINGO2) which
will have greater sensitivity (to allow observations of several bursts
per year) and time resolution ($\sim1$ second for bright bursts) and will
begin operations on the Liverpool Telescope in late 2009.

% For Figures insertion uncomment the next section
%%%%%%%%%%%%%%%%%%%%%%%%%%%%%%%%%%%%%%%%%%%%%%%%%%%%%%%%
% End of the paper
%
\end{document}